\documentclass[prd,twocolumn,groupedaddress,showpacs,nofootinbib]{revtex4-1}
\usepackage{graphicx}
\usepackage{dcolumn}
\usepackage{amssymb}
\usepackage{amsmath}
\usepackage{epsfig}
\usepackage[dvips]{color}
\usepackage{hhline}

\begin{document}

\title{ Leptogenesis with TeV Scale $W_R$}

\author{Pei-Hong Gu$^1$}
\email{peihong.gu@sjtu.edu.cn}

\author{Rabindra N. Mohapatra$^2$}
\email{rmohapat@umd.edu}

\affiliation{$^1$School of Physics and Astronomy, Shanghai Jiao Tong
University, 800 Dongchuan Road, Shanghai 200240, China\\
$^2$Maryland Center for Fundamental Physics and Department of Physics, 
University of Maryland, College Park, Maryland 20742}

\begin{abstract}
Successful leptogenesis within the conventional TeV-scale left-right implementation of type-I seesaw has been shown to require that the mass of the right-handed 
$W_R^\pm$ boson should have a lower bound much above the reach of the Large Hadron Collider. This bound arises from the necessity to suppress the washout of lepton asymmetry due to $W_R^\pm$-mediated $\Delta L\neq 0$ processes. We show that in an alternative quark seesaw realization of left-right symmetry, the above bound can be avoided. Lepton asymmetry in this model is generated not via the usual right-handed neutrino decay but rather  via the decay of new heavy scalars producing an asymmetry in the $B-L$ carrying Higgs triplets responsible for type-II seesaw, whose subsequent decay leads to the lepton asymmetry. This result implies that any evidence for $W_R$ at the LHC 14 will point towards this alternative realization of left-right symmetry, which is also known to solve the strong CP problem.

\end{abstract}


\maketitle

\section{Introduction}

The discoveries of atmospheric, solar, accelerator and reactor neutrino oscillations have established that the neutrinos are massive and are mixed among different flavors \cite{patrignani2016}. This is the first evidence for new physics beyond the standard $SU(3)_c^{}\times SU(2)_L^{}\times U(1)^{}_{Y}$  model (SM). Meanwhile, the cosmological observations combined with oscillation data imply that the neutrino masses should be in a sub-eV range \cite{patrignani2016}. A simple beyond the standard model (BSM) mechanism for understanding such tiny neutrino masses is provided by the various seesaw schemes \cite{minkowski1977,mw1980,mv1986,flhj1989}, known as type-I, type-II, inverse and type-III seesaw respectively. In the type I and inverse seesaw scenarios, the SM is extended by the inclusion of heavy SM singlet fermions, i.e. the right-handed neutrinos (RHNs) with Majorana masses (for type-I and inverse) and in type II case, heavy triplet scalars with lepton number violating couplings are added to SM. After the $SU(2)_L^{}\times U(1)^{}_{Y}$ electroweak symmetry is spontaneously broken down to the $U(1)_{em}^{}$, the neutrinos acquire a Majorana mass term highly suppressed by a small ratio of the electroweak scale over the heavy Majorana fermion or scalar masses. These models have the extra advantage that they provide a way to understand the cosmic matter-antimatter asymmetry via the so-called leptogenesis mechanism \cite{fy1986}. In the type-I seesaw case, the typical scenario involves the decay of heavy RHNs of the seesaw to generate a lepton asymmetry through their Yukawa interactions as long as CP is not conserved \cite{fy1986}. Subsequently, this lepton asymmetry is partially converted to a baryon asymmetry through the sphaleron processes \cite{krs1985}.

Left-right (LR) symmetric models based on the gauge group $SU(3)_c^{}\times SU(2)_L^{}\times SU(2)_R^{}\times U(1)_{B-L}$ \cite{ps1974} provide an elegant embedding of seesaw mechanism into more fundamental framework. It naturally provides the two key ingredients of the type-I seesaw mechanism i.e. the RHNs which arise from anomaly cancellation and the Majorana masses for RHNs which arise from the breaking of the $SU(2)_R^{}\times U(1)_{B-L}^{}$ symmetry. In the LR symmetric models, the SM left-handed fermions are placed in the $SU(2)_L^{}$ doublets as they are in the SM, the SM right-handed fermions plus three RHNs are placed in the $SU(2)_R^{}$ doublets. Implementation of type-I seesaw in the model comes from the Higgs scalars consisting of $SU(2)_{L,R}$ triplets and one $SU(2)_L^{}\times SU(2)_R^{}$ bidoublet. When the $[SU(2)_R^{}]$-triplet Higgs scalar develops its vacuum expectation value (VEV) breaking the LR symmetry down to the SM, i.e. $SU(2)^{}_{R} \times U(1)^{}_{B-L} \longrightarrow U(1)^{}_{Y}$ and the lepton number violation needed for seesaw emerges automatically. After the RHNs go out of equilibrium in the early universe, their decays then can provide a realization of leptogenesis. Because the RHNs couple to the non-SM gauge bosons in the LR model, their departure from equilibrium as well as the suppression of subsequent wash-out constrains the LR symmetry breaking \cite{fhv2009,dlm2015}, implying $M_{W_R^{}}\geq 10-13$ TeV depending on assumptions. This implies that it would be impossible to simultaneously have a successful leptogenesis and LR symmetry with type-I seesaw testable at the LHC.

In addition to the neutrino mass and the baryon asymmetry, the SM faces another challenge, the strong CP problem. It is an interesting property of the LR symmetric models that they provide an alternative setup to solve the strong CP problem without the need for an axion. This is because these models naturally contain a discrete parity symmetry which makes the tree-level quark mass matrices hermitian, implying the vanishing of the strong CP phase at the tree level \cite{ms1978}. This  leads to a calculable value for $\theta$ from loop effects. However in the minimal LR type-I seesaw model, additional symmetries (e.g. supersymmetry~\cite{susyLR}) are required for this purpose. A realization of left-right symmetry, where a finite $\theta$ is generated at the two-loop level without the need for any additional symmetries~\cite{bm1989} and is therefore naturally below the current limits, is provided by adding heavy vector-like quarks and leptons to the model. The quark masses arise in this model from a seesaw mechanism  (called quark seesaw \cite{berezhiani1983} in this paper) using the vector-like quark masses as the heavy end of the seesaw. The model has a different Higgs structure i.e. a left doublet and a right doublet for the corresponding $SU(2)$'s, instead of a bi-doublet and $B-L=2$ triplets as in the minimal LR type-I seesaw case. 

Our goal in this paper is to focus on this class of quark seesaw LR symmetric models and discuss leptogenesis and neutrino masses.  We extend the minimal quark seesaw LR model by adding $B-L=2$ Higgs triplets to make it realistic in the neutrino sector. The  small neutrino masses are generated using type-II seesaw mechanism. The required smallness of the left triplet VEV needed for this purpose arises from a two-step suppression of the VEVs ~\cite{ghsz2009} with no couplings smaller than $\sim 10^{-5}$ (which is roughly of the same order as the electron Yukawa in the SM). This is a significant improvement over the canonical type-II models in the SM where couplings as small as $10^{-10}$ or masses as large as $10^{12}_{}\,\textrm{GeV}$ are required. We keep all masses in our model below $100\textrm{TeV}$.

We then discuss leptogenesis in this model and show that the lower bound on the $W_R$ mass mentioned above for the minimal type-I LR model can be avoided, making $W_R$ accessible at the LHC. The way it comes about is that, the lepton asymmetry arises not from  the decay of RHNs but rather the decay of new heavy SM singlet scalar bosons, as for instance in~\cite{ghsz2009}. This heavy scalar decay leads to an asymmetry in $B-L$ carrying scalar triplets responsible for type-II seesaw. The triplet scalars decay to leptons generating the lepton asymmetry that gets converted to the baryon asymmetry via the sphaleron interactions. There are no $W_R$ mediated wash-out processes in this case, thus avoiding the bound on $W_R$ mass. The significance of this result is that if the $W_R$ is discovered at the LHC 14, it will not only provide evidence for the idea of LR extension of SM but also point towards the quark seesaw realization of the idea as well as providing possibly alternative solution to the strong CP problem. This will have significant implication for physics beyond the SM e.g. any possible grand unification, proton decay etc. The model also has the feature that the right handed neutrino masses are in the keV range, with implications for low energy beta decay phenomenology~\cite{werner}.

The paper is organized as follows: in sec. II, we review the quark seesaw realization of the left-right model with new Higgs scalar triplets for neutrino masses and the different symmetry breaking stages; in sec III, we discuss how small neutrino masses arise in this model; In sec. IV, we give a benchmark values for parameters of the model; in sec. V, we present the calculation of the lepton asymmetry generation and in sec. Vi we discuss how this lepton asymmetry gets converted to the baryon symmetry. and in sec, VII, we discuss some phenomenological implications and in sec. VIII, we summarize our results.

\section{The model}

Our model includes the chiral fermions and Higgs scalars as given below,
\begin{eqnarray}
\label{fermion1}
&&\begin{array}{c}q^{}_{L}(3,2,1,+\frac{1}{3})(+\frac{1}{3})\end{array}=\left[\begin{array}{c}u^{}_{L}\\
[2mm]
d^{}_{L}\end{array}\right]\,,\nonumber\\
[2mm]
&&\begin{array}{c}q^{}_{R}(3,1,2,+\frac{1}{3})(+\frac{1}{3})\end{array}=\left[\begin{array}{c}u^{}_{R}\\
[2mm]
d^{}_{R}\end{array}\right]\,,\nonumber\\
[2mm]
&&\begin{array}{c}D_{L,R}^{}(3,1,1,-\frac{2}{3})(+\frac{1}{3})\,,~~U_{L,R}^{}(3,1,1,+\frac{4}{3})(+\frac{1}{3})\,;\end{array}\nonumber\\
[2mm]
&&\begin{array}{c}l^{}_{L}(1,2,1,-1)(-1)\end{array}=\left[\begin{array}{c}\nu^{}_{L}\\
[2mm]
e^{}_{L}\end{array}\right]\,,\nonumber\\
[2mm]
&&\begin{array}{c}l^{}_{R}(1,1,2,-1)(-1)\end{array}=\left[\begin{array}{c}\nu^{}_{R}\\
[2mm]
e^{}_{R}\end{array}\right]\,,\nonumber\\
[2mm]
&&E_{L,R}^{}(1,1,1,-2)(-1)\,;\nonumber\\
[2mm]
&&\begin{array}{c}\phi_{L}^{}(1,2,1,-1)(0)\end{array}=\left[\begin{array}{c}\phi^{0}_{L}\\
[2mm]
\phi^{-}_{L}\end{array}\right]\,,\nonumber\\
[2mm]
&&\begin{array}{c}\phi_{R}^{}(1,1,2,-1)(0)\end{array}=\left[\begin{array}{c}\phi^{0}_{R}\\
[2mm]
\phi^{-}_{R}\end{array}\right]\,,\nonumber\\
[2mm]
&&\begin{array}{c}\Delta_{L}^{}(1,3,1,+2)(+2)\end{array}=\left[\begin{array}{ll}\delta^{+}_{L}/\sqrt{2}& \delta^{++}_{L}\\
[2mm]
\delta^{0}_{L}& \delta^{+}_{L}/\sqrt{2}\end{array}\right]\,,\nonumber\\
[2mm]
&&\begin{array}{c}\Delta_{R}^{}(1,1,3,+2)(+2)\end{array}=\left[\begin{array}{ll}\delta^{+}_{R}/\sqrt{2}& \delta^{++}_{R}\\
[2mm]
\delta^{0}_{R}& \delta^{+}_{R}/\sqrt{2}\end{array}\right]\,;\nonumber\\
[2mm]
&&\sigma(1,1,1,0)(-2)\,,~~\chi(1,1,1,0)(+1)\,.
\end{eqnarray}
Here the first brackets following the fields describe the transformations under the $G_{LR}^{}=SU(3)_c^{}\times SU(2)_L^{}\times SU(2)_R^{}\times U(1)_{X}^{}$ gauge group, while the second brackets are the global $B-L$ numbers\footnote{The present $U(1)_{B-L}^{}$ global symmetry can be a gauged symmetry. But we do not discuss it here.}. For brevity, we do not write down the full Lagrangian. Instead we only show the following terms relevant for our discussion.
\begin{eqnarray}
\label{lag}
\mathcal{L}&\supset&-y_D^{}\left(\bar{q}_L^{}\tilde{\phi}_L^{}D_R^{}+\bar{q}_R^{}\tilde{\phi}_R^{}D_L^{}\right)-M_D^{}\bar{D}_R^{}D_L^{}\nonumber\\
[2mm]
&&-y_U^{}\left(\bar{q}_L^{}\phi_L^{}U_R^{}+\bar{q}_R^{}\phi_R^{}U_L^{}\right)-M_U^{}\bar{U}_R^{}U_L^{}\nonumber\\
[2mm]
&&-y_E^{}\left(\bar{l}_L^{}\tilde{\phi}_L^{}E_R^{}+\bar{l}_R^{}\tilde{\phi}_R^{}E_L^{}\right)-M_E^{}\bar{E}_R^{}E_L^{}\nonumber\\
[2mm]
&&-\frac{1}{2}f\left(\bar{l}_L^c i \tau_2^{} \Delta_L^{} l_L^{}+\bar{l}_R^c i \tau_2^{} \Delta_R^{} l_R^{}\right)\nonumber\\
[2mm]
&&-\kappa \sigma \left(\phi_L^T i\tau_2^{} \Delta_L^{} \phi_L^{}+\phi_R^T i\tau_2^{} \Delta_R^{} \phi_R^{}\right)-\mu \sigma \chi^2_{}\nonumber\\
[2mm]
&&+\textrm{H.c.}+M_{\Delta_L}^2 \textrm{Tr}\left(\Delta_L^\dagger \Delta_L^{}\right)+M_{\Delta_R}^2 \textrm{Tr}\left(\Delta_R^\dagger \Delta_R^{}\right)\nonumber\\
[2mm]
&&+M_{\sigma}^2 \sigma^\dagger_{}\sigma \,.
\end{eqnarray}

The $[SU(2)]$-doublet Higgs scalars $\phi^{}_{L,R}$ are responsible for the left-right and electroweak symmetry breaking, i.e. 
\begin{eqnarray}
&&SU(2)_L^{}\times SU(2)_R^{}\times U(1)_{X}^{} \nonumber\\
&&\stackrel{\langle\phi_R^{}\rangle}{\longrightarrow} SU(2)_L^{}\times U(1)_{Y}^{} \stackrel{\langle\phi_L^{}\rangle}{\longrightarrow}  U(1)_{em}^{} \,.
\end{eqnarray}
This can be achieved by softly breaking the parity symmetry in Eq. (\ref{lag}) in the Higgs doublet masses. This keeps the strong CP solution unaffected.

Note that at this stage the gauge symmetry is fully broken down to $SU(3)_c\times U(1)_{em}$ as desired but it leaves a global $B-L$ symmetry unbroken. Thus the neutrinos only have a tiny Dirac mass from two-loop diagram~\cite{chang1987}. When the $[G_{LR}^{}]$-singlet Higgs scalar $\chi$ develops its VEV $\langle\chi\rangle$ it breaks the global $U(1)_{B-L}^{}$ symmetry. We assume the scale of this breaking to be $\sim 100$ TeV. The corresponding massless Goldstone boson decouples from the thermal bath at a high temperature and does not affect teh subsequent evolution of the universe. The other $[G_{LR}^{}]$-singlet Higgs scalars $\sigma$ then acquires a seesaw-suppressed VEV, 
\begin{eqnarray}
\label{vevs}
\langle\sigma\rangle\simeq -\frac{\mu\langle\chi\rangle^2_{}}{M_{\sigma}^2}\ll \langle\chi\rangle ~~\textrm{for}~~M_{\sigma}^{2}\gg  \mu\langle\chi\rangle\,.
\end{eqnarray}
As for the $[SU(2)]$-triplet Higgs scalars $\Delta_{L,R}^{}$, their VEVs arise from the $\sigma$ vev and are therefore highly suppressed by the small VEVs $\langle\sigma\rangle$ as well as by the small value of the coupling parameter $\kappa$ which is required for generation of matter-anti-matter asymmetry(see later), i.e.
\begin{eqnarray}
\label{vevd}
\langle\Delta_{L,R}^{}\rangle\simeq -\frac{\kappa\langle\sigma\rangle\langle\phi_{L,R}^{}\rangle^2_{}}{M_{\Delta_{L,R}^{}}^2}\ll \langle\phi_{L,R}^{}\rangle\,.
\end{eqnarray} 
This for example leads to a VEV of $\Delta_L$ in the eV range, which the explains the tiny neutrino masses.

We emphasize that the phase in the cubic coupling $\kappa \langle\sigma\rangle$ can contribute an overall phase to the mass matrix $M_E^{}$ by redefining the fields $(\Delta_{L}^{},l_{L}^{},E_R^{})$ and $(\Delta_R^{},l_{R}^{},E_{L}^{})$. This overall phase can not be removed because the parity symmetry now is softly broken but its effect on the strong CP phase arises in high loop level and the model still solves the strong CP problem..

\section{Fermion masses}

After the left-right and electroweak symmetry breaking, we obtain the charged fermion masses,
\begin{eqnarray}
\mathcal{L}&\supset&-\left[\begin{array}{ll}\bar{u}_L^{}&\bar{U}_L^{}\end{array}\right]
\left[\begin{array}{cc}0&y_U^{}\langle\phi_L^{}\rangle\\
[2mm]y_U^{\dagger}\langle\phi_R^{}\rangle&M_U^{\dagger}\end{array}\right]
\left[\begin{array}{c}u_R^{}\\
[2mm]U_R^{}\end{array}\right]\nonumber\\
[2mm]
&&-\left[\begin{array}{ll}\bar{d}_L^{}&\bar{D}_L^{}\end{array}\right]
\left[\begin{array}{cc}0&y_D^{}\langle\phi_L^{}\rangle\\
[2mm]y_D^{\dagger}\langle\phi_R^{}\rangle&M_D^{\dagger}\end{array}\right]
\left[\begin{array}{c}d_R^{}\\
[2mm]D_R^{}\end{array}\right]\nonumber\\
[2mm]
&&-\left[\begin{array}{ll}\bar{e}_L^{}&\bar{E}_L^{}\end{array}\right]
\left[\begin{array}{cc}0&y_E^{}\langle\phi_L^{}\rangle\\
[2mm]y_E^{\dagger}\langle\phi_R^{}\rangle&M_E^{\dagger}\end{array}\right]
\left[\begin{array}{c}e_R^{}\\
[2mm]E_R^{}\end{array}\right]+\textrm{H.c.}\,.\nonumber\\
&&
\end{eqnarray}
It is well known the above quark mass matrices can solve the strong CP problem without introducing the axion. The details can be found in \cite{bm1989}. Furthermore, since the $[SU(2)]$-singlet fermions are heavy, they can be integrated out. The SM fermion masses thus should be 
\begin{eqnarray}
m_d^{}&=&-y_D^{}\frac{\langle\phi_L^{}\rangle\langle\phi_R^{}\rangle}{M_D^{}}y_D^\dagger\,,\nonumber\\
m_u^{}&=&-y_U^{}\frac{\langle\phi_L^{}\rangle\langle\phi_R^{}\rangle}{M_U^{}}y_U^\dagger\,,\nonumber\\
m_e^{}&=&-y_E^{}\frac{\langle\phi_L^{}\rangle\langle\phi_R^{}\rangle}{M_E^{}}y_E^\dagger\,.
\end{eqnarray}

We now study the neutrino mass. Clearly, through the Yukawa couplings of the $[SU(2)]$-triplet Higgs scalars $\Delta_{L,R}^{}$ to the $[SU(2)]$-doublet leptons $l_{L,R}^{}$, the left- and right-handed neutrinos can obtain their Majorana masses, 
\begin{eqnarray}
\mathcal{L}&\supset&-\frac{1}{2} m_L^{}\bar{\nu}_L^c \nu_L^{}  -\frac{1}{2} m_R^{}\bar{\nu}_R^c \nu_R^{} +\textrm{H.c.}\nonumber\\
&&\textrm{with} ~~m_{L,R}^{}=f \langle\Delta_{L,R}^{}\rangle\,.
\end{eqnarray}
Thanks to the small VEV $\langle\Delta_{L}^{}\rangle$, the left-handed neutrino masses $m_{L}^{}$ can be made tiny even if the Yukawa couplings $f$ are sizable. This is the essence of the type-II seesaw. As shown in Eq. (\ref{vevs}), the present model has a first stage seesaw suppression of the parameter $\kappa\langle\sigma\rangle$ in a natural way and a second step suppression due to the small coupling $]kappa\sim 10^{-5}$, which is only of the same order as eelctron Yukawa coupling in the SM  and hence not overly fine-tuned.  Clearly, the same two step suppressed type-II seesaw mechanism also applies to the generation of the right-handed neutrino masses $m_R^{}$. As a result, the right-handed neutrino masses are much below the left-right symmetry breaking scale and for our benchmark choice of parameters (sec.VI) are in few keV range. A cosmologically relevant observation is that since $\langle\Delta_{L,R}^{}\rangle$ is proportional to $\langle\sigma\rangle\sim $ GeV, the neutrinos are massless until the universe cools down to $T\sim 0.1-1$ GeV and are massless when the sphaleron interactions are in equilibrium.
 
As noted, at the two-loop level, the charged current interactions and the charged fermion masses can always mediate a Dirac mass term between the left- and right-handed neutrinos \cite{chang1987}\footnote{ Because of some quartic couplings in the scalar potential, the singly charged components of the $[SU(2)]$-triplet Higgs scalars $\Delta_{L}^{}$ and $\Delta_{R}^{}$ can mix with each other at tree level after the left-right and electroweak symmetry breaking. These singly charged scalars with the charged leptons can mediate a one-loop diagram to generate a Dirac mass term between the left- and right-handed neutrinos. This Dirac mass term could be much smaller than the two-loop contribution since the VEVs $\langle\Delta_{L,R}^{}\rangle$ are assumed much smaller than the quark masses $m_{t,b}^{}$.}, 
\begin{eqnarray}
\label{dnumass}
\mathcal{L}&\supset&- m_D^{}\bar{\nu}_L^{}\nu_R^{}+\textrm{H.c.}~~\textrm{with} ~~m_D^{}\simeq \frac{3g^4_{}}{4(16\pi^2_{})^2_{}}\frac{m_t^{}m_b^{}}{M_{W_R^{}}^2}\hat{m}_e\,.\nonumber\\
&&
\end{eqnarray}
The $\nu_L^{}-\nu_R^{}$ mixing is small enough enough to escape from the BBN constraint~\cite{bbn}. 

\section{Numerical example}

In this section, we give a benchmark scenario to show how the model works for right handed $W_R$ mass in the low TeV range so that it is accessible at the LHC. For example, we take  
\begin{eqnarray}
\label{par1}
\langle\phi_R^{}\rangle =5\,\textrm{TeV}\Longrightarrow M_{W_R^{}}^{}=\frac{1}{\sqrt{2}}g \langle\phi_R^{}\rangle \simeq 2.3\,\textrm{TeV}\,.
\end{eqnarray} 
In addition, we assume the global lepton number breaking scale to be\footnote{If the present $U(1)_{B-L}^{}$ global symmetry is a gauged symmetry, we need consider the departure from equilibrium of the related gauge interactions. Furthermore, we need check other experimental constraints \cite{cddt2004}. For this paper, we keep it as a global symmetry.}
\begin{eqnarray}
\label{par2}
\langle\chi\rangle =100\,\textrm{TeV}\,.
\end{eqnarray}
By inserting
\begin{eqnarray}
\label{par3}
&&M_{\Delta_{L}^{}}^{}=3\,\textrm{TeV}\,,~~M_{\Delta_{R}^{}}^{}=1\,\textrm{TeV}\,,~~M_{\sigma_{1,2}^{}}^{}=100\,\textrm{TeV}\,,\nonumber\\
&&\mu_{1,2}^{}=10^{-6}_{}\,M_{\sigma_{1,2}^{}}^{}\,,~~\kappa_{1,2}^{}=10^{-5}_{}\,,
\end{eqnarray}
in Eqs. (\ref{vevs}) and (\ref{vevd}), we then read
\begin{eqnarray}
\label{par4}
\langle\sigma\rangle\simeq 100\,\textrm{MeV}\,,~~\langle\Delta_{R}^{}\rangle\simeq 25\,\textrm{keV}\,,~~\langle\Delta_{L}^{}\rangle\simeq 3\,\textrm{eV}\,.
\end{eqnarray}
Eventually, the neutrino masses should be 
\begin{eqnarray}
\label{par5}
&&m_L^{}= 3\,\textrm{eV}\cdot f_\Delta^{} \,,~~m_R^{}=25\,\textrm{keV}\cdot f_\Delta^{}\,,\nonumber\\
&&m_D^{}\simeq 1.3\,\textrm{eV}\,,
\end{eqnarray}
Note the right-handed neutrinos $\nu_R^{}$ can be quasi-degenerate even if the left-handed neutrinos $\nu_L^{}$ have a hierarchical spectrum. This is because the parity symmetry is softly broken and hence the right-handed PMNS matrix is allowed different from the left-handed PMNS matrix. Accordingly, the three $\nu_R^{}$ can be all at the keV scale while the $\nu_L^{}-\nu_R^{}$ mixing $\theta_{LR}^{} \sim m_D^{}/m_R^{}$ can be very small when the Yukawa couplings $f_\Delta^{}$ are sizable, i.e. 
\begin{eqnarray}
\label{par6}
\theta_{LR}^{} \sim m_D^{}/m_R^{}\lesssim 10^{-3}_{}~~\textrm{for}~~f_\Delta^{}=0.05\,.
\end{eqnarray}

We now show that with our choice of parameters, the keV right-handed neutrinos do affect the BBN discussion. First, the right-handed neutrinos will decouple from the weak interactions at the temperature around $T\sim 3\,\textrm{MeV}\, \theta_{LR}^{-2/3}\simeq 300\,\textrm{MeV}(\theta_{LR}^{}/10^{-3}_{})^{-2/3}_{}$ where $3\, \textrm{MeV}$ is the decoupling temperature for the left-handed neutrinos. Secondly, the annihilating and scattering processes mediated by the other gauge bosons will decouple at the temperature around $T\sim 3\,\textrm{MeV} (M_{W_R}^{}/M_{W_L}^{})^{4/3}_{}\simeq 260\,\textrm{MeV}(M_{W_R}^{}/2.3\,\textrm{TeV})^{4/3}_{}$. Thirdly, we consider the annihilating and scattering processes mediated by the singly charged component of the $\Delta_R^{}$ scalar. The decouple temperature should be about $T\sim 3\,\textrm{MeV} (g/f_\Delta^{})^{4/3}_{} (M_{\Delta_R}^{}/M_{W_L}^{})^{4/3}_{}\simeq 2.7\,\textrm{GeV} (0.05/f_\Delta^{})^{4/3}_{}(M_{\Delta_R}^{}/1\,\textrm{TeV})^{4/3}_{}$. In conclusion, the existence of the keV right-handed neutrinos will not conflict with the BBN. As already noted, the $\nu_L-\nu_R$ mixing is small enough so that oscillations do not affect the BBN considerations~\cite{bbn}.

\section{Lepton asymmetry}

We now discuss the generation of lepton asymmetry in our model. As noted, it is different from the mechanism in type I seesaw models where decay of the right handed neutrinos was the main source. In our model, it is the decay of heavy singlet scalar $\sigma$ decay which plays this role.
Before the left-right symmetry breaking, the $[G_{LR}^{}]$-singlet scalars $\sigma_a^{}$ can have three decay modes, $\sigma_a^{}\rightarrow \phi^\ast_{L}\phi^\ast_{L}\Delta_{L}^\ast $, $\sigma_a^{}\rightarrow \phi^\ast_{R}\phi^\ast_{R}\Delta_{R}^\ast $ and $\sigma_a^{}\rightarrow \chi^\ast_{}\chi^\ast_{}$. We calculate the decay width at tree level,
\begin{eqnarray}
\label{width}
\Gamma_{\sigma_a^{}}^{}&=&\Gamma(\sigma_a^{}\longrightarrow \phi^\ast_{L}\phi^\ast_{L}\Delta_{L}^\ast )+\Gamma(\sigma_a^{}\longrightarrow \phi^\ast_{R}\phi^\ast_{R}\Delta_{R}^\ast )\nonumber\\
[2mm]
&&+\Gamma(\sigma_a^{}\longrightarrow \chi^\ast_{}\chi^\ast_{})\nonumber\\
[2mm]
&\equiv&\Gamma(\sigma_a^{\ast}\longrightarrow \phi^{}_{L}\phi^{}_{L}\Delta_{L}^{})+\Gamma(\sigma_a^{\ast}\longrightarrow \phi^{}_{R}\phi^{}_{R}\Delta_{R}^{} )\nonumber\\
[2mm]
&&+\Gamma(\sigma_a^{\ast}\longrightarrow \chi \chi )\,,\nonumber\\
&=&\frac{1}{8\pi}\left(\frac{|\mu_a^{}|^2}{M_{\sigma_a^{}}^2}+\frac{3|\kappa_a^{}|^2}{16\pi^2_{}}\right)M_{\sigma_a^{}}^{}\,.
\end{eqnarray}
We also compute the one-loop CP asymmetries in the above decays, i.e.
\begin{eqnarray}
\label{cpa}
\varepsilon_{\sigma_a^{}}^{\Delta_L^{}}&=&2\frac{\Gamma(\sigma_a^{}\longrightarrow \phi^\ast_{L}\phi^\ast_{L}\Delta_{L}^\ast )-\Gamma(\sigma_a^{\ast}\longrightarrow \phi^{}_{L}\phi^{}_{L}\Delta_{L}^{} )}{\Gamma_{\sigma_a^{}}^{}}\nonumber\\
&=&\frac{1}{4\pi}\frac{\textrm{Im}\left(\mu_b^\ast\mu_a^{}\kappa_b^{}\kappa_a^\ast\right)}{M_{\sigma_{a}^{}}^2-M_{\sigma_b^{}}^2}\frac{1}{\frac{|\mu_a^{}|^2}{M_{\sigma_a^{}}^2}+\frac{3|\kappa_a^{}|^2}{16\pi^2_{}}}\,,\nonumber\\
[2mm]
\varepsilon_{\sigma_a^{}}^{\Delta_R^{}}&=&2\frac{\Gamma(\sigma_a^{}\longrightarrow \phi^\ast_{R}\phi^\ast_{R}\Delta_{R}^\ast )-\Gamma(\sigma_a^{\ast}\longrightarrow \phi^{}_{R}\phi^{}_{R}\Delta_{R}^{} )}{\Gamma_{\sigma_a^{}}^{}}\nonumber\\
&=&\frac{1}{4\pi}\frac{\textrm{Im}\left(\mu_b^\ast\mu_a^{}\kappa_b^{}\kappa_a^\ast\right)}{M_{\sigma_{a}^{}}^2-M_{\sigma_b^{}}^2}\frac{1}{\frac{|\mu_a^{}|^2}{M_{\sigma_a^{}}^2}+\frac{3|\kappa_a^{}|^2}{16\pi^2_{}}}\,,\nonumber\\
[2mm]\varepsilon_{\sigma_a^{}}^{\chi}&=&\frac{\Gamma(\sigma_a^{}\longrightarrow \chi^\ast_{}\chi^\ast_{})-\Gamma(\sigma_a^{\ast}\longrightarrow \chi \chi)}{\Gamma_{\sigma_a^{}}^{}}\nonumber\\
&=&-\frac{1}{4\pi}\frac{\textrm{Im}\left(\mu_b^\ast\mu_a^{}\kappa_b^{}\kappa_a^\ast\right)}{M_{\sigma_{a}^{}}^2-M_{\sigma_b^{}}^2}\frac{1}{\frac{|\mu_a^{}|^2}{M_{\sigma_a^{}}^2}+\frac{3|\kappa_a^{}|^2}{16\pi^2_{}}}\,.\end{eqnarray}
It is easy to check that 
\begin{eqnarray}
\varepsilon_{\sigma_a^{}}^{\Delta_L^{}}=\varepsilon_{\sigma_a^{}}^{\Delta_R^{}}\equiv \frac{1}{2}\varepsilon_{\sigma_a^{}}^{\Delta}\,,
\end{eqnarray}
due to the parity symmetry, meanwhile,
\begin{eqnarray}
\varepsilon_{\sigma_a^{}}^{\Delta_L^{}}+\varepsilon_{\sigma_a^{}}^{\Delta_R^{}}+\varepsilon_{\sigma_a^{}}^{\chi}\equiv 0\,,
\end{eqnarray}
due to the lepton number conservation.

When the $[G_{LR}^{}]$-singlet scalars $\sigma$ go out of equilibrium, their decays can produce a lepton asymmetry stored in the $[SU(2)]$-triplet Higgs scalars $\Delta_{L,R}^{}$ and an opposite lepton asymmetry stored in the $[G_{LR}^{}]$-singlet Higgs scalar $\chi$. Clearly, the $\Delta_{L,R}^{}$ asymmetry and the $\chi$ asymmetry decouple from each other once they are induced. This is clearly satisfied in our model since the communication between the $\chi$ - field and the $\Delta_{L,R}$ field only occurs via the $\sigma$ mixing with $\chi$'s from $\mu$ coupling and $\kappa$ coupling after $\langle\phi_R\rangle$ switches on. This mixing is given by the product of two small parameters $\mu$ and $\kappa$ in our choice of parameters and is of order $\sim 10^{-11}$. Thus this mixing is not cosmologically relevant at the epoch of  leptogenesis.  The $[SU(2)]$-triplet Higgs scalars $\Delta_{L,R}^{}$ can transfer their lepton asymmetry to the $[SU(2)]$-doublet leptons $l_{L,R}^{}$ when they decay and this happens before the sphaleron processes become inactive at temperature $T_{\textrm{sph}}^{}\sim 100\,\textrm{GeV}$.

In this model, the spontaneous breaking of the global $U(1)_{B-L}^{}$ symmetry occurs around 200 TeV by the $\chi$ VEV; however due to low value for $\langle\sigma\rangle$, no lepton-number-violating processes would emerge before the sphalerons had partially converted the lepton asymmetry to a baryon asymmetry.
A consequence of this is that  the lepton-number-violating processes $l_{L(R)}^{}l_{L(R)}^{}\longleftrightarrow \phi^{}_{L(R)}\phi^{}_{L(R)}$, $l_{L(R)}^{c}  l_{L(R)}^{c}\longleftrightarrow \phi^{\ast}_{L(R)} \phi^{\ast}_{L(R)}$, $l_{L(R)}^{}\phi_{L(R)}^{\ast} \longleftrightarrow l_{L(R)}^{c} \phi^{}_{L(R)}$, which could have a potential wash out effect on the lepton asymmetry do not go into equilibrium before the sphalerons stop working, and therefore do not pose a problem. To be more explicit, we demand that
\begin{eqnarray}
\label{con}
\Gamma_{\Delta L=2}^{}<  H(T)~~\textrm{for}~~T>T_{\textrm{sph}}^{}\,.
\end{eqnarray}
where the Hubble constant $H(T)$ is given by
\begin{eqnarray}
\label{hubble}
H(T)=\left(\frac{8\pi^{3}_{}g_{\ast}^{}}{90}\right)^{\frac{1}{2}}_{}
\frac{T^{2}_{}}{M_{\textrm{Pl}}^{}}\,,
\end{eqnarray}
with $M_{\textrm{Pl}}^{}\simeq 1.22\times 10^{19}_{}\,\textrm{GeV}$ being the Planck mass and $g_{\ast}^{}=\mathcal{O}(100)$ being the relativistic degrees of freedom. As for the lepton-number-violating interaction rates $\Gamma_{\Delta L=2}^{}$, they can be estimated by 
\begin{eqnarray}
\Gamma_{\Delta L=2}^{}&\sim& \frac{\textrm{Tr}(f^\dagger_{}f)|\kappa_a^{}\langle\sigma_a^{}\rangle|^2_{}}{T} ~~~~~\textrm{for}~~T>M_{\Delta_{L,R}^{}}^{}\,,\nonumber\\
[5mm]
\Gamma_{\Delta L=2}^{}&\sim& \frac{\textrm{Tr}(f^\dagger_{}f)|\kappa_a^{}\langle\sigma_a^{}\rangle|^2_{}T^3_{}}{M_{\Delta_{L,R}^{}}^{4}} ~~\textrm{for}~~T<M_{\Delta_{L,R}^{}}^{}\,.
\end{eqnarray}
Eq. (20) is satisfied for these processes for our choice of  $\Delta_L$ mass of 3 TeV.

We need also check the right-handed neutrino-antineutrino oscillation, which breaks lepton number by two units is out of equilibrium till below the sphaleron decoupling temperature. It is well known this type of $\nu_R^{}-\bar{\nu}_R^{}$ oscillation should be helicity suppressed \cite{bp1978}. We can roughly estimate the rate by
\begin{eqnarray}
\Gamma_{\nu_R^{}-\bar{\nu}_R^{}}^{}\sim \frac{\textrm{Tr}\left(m_R^\dagger m_R^{}\right)}{T}~~\textrm{for}~~T\gg m_R^{}\,.
\end{eqnarray}
If the right-handed charged current interactions,
\begin{eqnarray}
\label{rcurrent}
\mathcal{L}\supset \left(\frac{M_{W_L}^2}{M_{W_R}^2}\right)\frac{G_F^{}}{\sqrt{2}} \bar{u}_R^{}\gamma^\mu_{}d_R^{} \bar{e}_R^{}\gamma_\mu^{}\nu_R^{}
+\textrm{H.c.}\,,
\end{eqnarray}
keep in equilibrium above the crucial temperature $T_{\textrm{sph}}^{}$, the lepton-number-violating $\nu_R^{}-\bar{\nu}_R^{}$ oscillation should match the following condition,
\begin{eqnarray}
\label{nunubar}
\Gamma_{\nu_R^{}-\bar{\nu}_R^{}}^{}< H(T)~~\textrm{for}~~T>T_{\textrm{sph}}^{}\,,
\end{eqnarray}
to avoid the elimination of the produced lepton asymmetry. It is easy to see that for keV right handed neutrinos, the above conditions are easily satisfied.

The $[SU(2)]$-triplet Higgs scalars $\Delta_{L,R}^{}$, which have become non-relativistic above the the temperature $T_{\textrm{sph}}^{}$, can efficiently decay into the $[SU(2)]$-doublet leptons $l_{L,R}^{}$ and  convert their asymmetry into lepton number asymmetry of the universe. There are no lepton number wash out process active above the $T_{\textrm{sph}}$, as just demonstrated  and the lepton number of course gets converted to baryon asymmetry via sphaleron interactions.

\section{Baryon asymmetry}

We can calculate the final baryon asymmetry in our model from the initial lepton asymmetry using the chemical potential identities \cite{ht1990}. Our considerations are slightly different from that of Ref. \cite{ht1990} since we also have right handed neutrinos interactions in equilibrium. For this purpose, we denote $\mu_{q,d,u,l,e,\nu,\phi}^{}$ as the chemical potentials of the SM fermions $q_{L}^{}(3,2,+1/6)$, $d_{R}^{}(3,1,-1/3)$, $u_{R}^{}(3,1,+2/3)$, $l_L^{}(1,2,-1/2)$, $e_R^{}(1,1,-1)$, the right-handed neutrinos $\nu_R^{}(1,0,0)$, and the SM Higgs scalar $\varphi(1,2,+1/2)$ respectively. Here the brackets following the fields describe the transformations under the $SU(3)_c^{}\times SU(2)_L^{}\times U(1)^{}_{Y}$ gauge groups. Above the electroweak scale, the SM Yukawa interactions yield the following relations,
\begin{eqnarray}
\label{smych}
&&-\mu_{q}^{}+\mu_{\varphi}^{}+\mu_{d}^{}=0\,,~~
 -\mu_{q}^{}-\mu_{\varphi}^{}+\mu_{u}^{}=0\,,\nonumber\\
 &&
 -\mu_{l}^{}+\mu_{\varphi}^{}+\mu_{e}^{}=0\,,
\end{eqnarray}
whereas the $SU(2)_L^{}$ sphalerons being in equilibrium lead to,
\begin{eqnarray}
\label{sphch}
3\mu_{q}^{}+\mu_{l}^{}&=&0\,,
\end{eqnarray}
and the vanishing hypercharge in the universe requires,
\begin{eqnarray}
\label{hyperc}
3\left(\mu_{q}^{}-\mu_{d}^{}+2\mu_{u}^{}-\mu_{l}^{}-\mu_{e}^{}\right)+2\mu_{\varphi}^{} =0\,.
\end{eqnarray}
At this stage, the right-handed charged current interactions (\ref{rcurrent}) may still keep in equilibrium, depending on the left-right symmetry breaking scale and we assume that they are, In which case, we further have
\begin{eqnarray}
\label{rightnu}
-\mu_{u}^{}+\mu_{d}^{}-\mu_e^{}+\mu_\nu^{}&=&0\,.
\end{eqnarray} 
Solving for these equations, all chemical potentials can be expressed in terms of a single one (chosen here as $\mu_{l}$. For example, we read
\begin{eqnarray}
&&\mu_q^{}=-\frac{1}{3}\mu_l^{}\,,~~\mu_d^{}=-\frac{19}{21}\mu_l^{}\,,~\mu_u^{}=\frac{5}{21}\mu_l^{}\,,~~\mu_e^{}=\frac{3}{7}\mu_l^{}\,,\nonumber\\
&&\mu_\nu^{}=\frac{11}{7}\mu_l^{}\,,~~\mu_\varphi^{}=\frac{4}{7}\mu_l^{}\,.
\end{eqnarray} 
The corresponding baryon and lepton asymmetries then should be 
\begin{eqnarray}
\label{relation}
B&=&3\left(2\mu_q^{}+\mu_u^{}+\mu_d^{}\right)=-4\mu_l^{}\,,\nonumber\\
L_{\textrm{SM}}^{}&=&3\left(2\mu_l^{}+\mu_e^{}\right)=\frac{51}{7}\mu_l^{}\,,\nonumber\\
L_{\nu_R}^{}&=&3\mu_\nu^{}=\frac{33}{7}\mu_l^{}\,.
\end{eqnarray}

If the right-handed charged current interactions (\ref{rcurrent}) have not decoupled before the $SU(2)_L^{}$ sphaleron processes stop working, we can easily read the final baryon asymmetry from the initial lepton asymmetries,
\begin{eqnarray}
\label{result1}
B^f_{}=\frac{1}{4}\left[B^i_{}-\left(L_{\textrm{SM}}^{i}+L_{\nu_R}^{i}\right)\right]= -\frac{1}{4}\left(L_{\textrm{SM}}^{i}+L_{\nu_R}^{i}\right)\,,
\end{eqnarray} 
by inserting the relation (\ref{relation}) and assuming the initial baryon asymmetry to be zero, i.e. $B^i_{}=0$.

We now demonstrate the generation of the lepton and baryon asymmetries. For the masses and couplings in Eq. (\ref{par3}), we can expect a weak washout condition, 
\begin{eqnarray}
K=\frac{\Gamma_{\sigma_{1,2}^{}}^{}}{H(T)}\left|_{T=M_{\sigma_{1,2}^{}}^{}}^{}\right.\simeq 0.6\,,
\end{eqnarray}
under which the final baryon asymmetry can well approximate to \cite{kt1990}
\begin{eqnarray}
\frac{n_B^{}}{s}=-\frac{1}{4}\frac{n_L^{}}{s}\sim -\frac{1}{4}\frac{\varepsilon_{\sigma_1^{}}^{\Delta}+\varepsilon_{\sigma_2^{}}^{\Delta}}{g_\ast^{}}\,.
\end{eqnarray}
Here $n_B^{}$, $n_L^{}$ and $s$ respectively are the baryon number density, the lepton number density and the entropy density, while $g_\ast^{}=209.5$ is the  the relativistic degrees of freedom during this leptogenesis epoch. Moreover, the conditions (\ref{con}) and (\ref{nunubar}) can be also satisfied. We further assume the $\sigma_{1,2}^{}$ scalars are quasi-degenerate, 
\begin{eqnarray}
\label{par7}
M_{\sigma_{2}^{}}^{}-M_{\sigma_{1}^{}}^{}=10^{-5}_{}\,M_{\sigma_{1,2}^{}}^{}\gg \Gamma_{\sigma_{1,2}^{}}^{}\,.
\end{eqnarray}
The CP asymmetries $\varepsilon_{\sigma_{1,2}^{}}^{\Delta}$ then can arrive at 
\begin{eqnarray}
\varepsilon_{\sigma_{1}^{}}^{\Delta}&=&\varepsilon_{\sigma_{2}^{}}^{\Delta}=2.74\times 10^{-7}_{}\,\sin\delta\nonumber\\
&&\textrm{with}~~\delta =\textrm{Arg}\left(\mu_2^\ast\mu_1^{}\kappa_2^{}\kappa_1^\ast\right)\,.
\end{eqnarray}
Therefore, the final baryon asymmetry can be consistent with the observed value,
\begin{eqnarray}
\frac{n_B^{}}{s}=10^{-10}_{}\left(\frac{\sin\delta}{0.15}\right)\,.
\end{eqnarray}

\section{Discussions and comments}

In this section, we make a few phenomenological comments on the model.
\begin{itemize}
\item The left-right symmetry breaking, the quark seesaw and the type-II seesaw scales are all at the TeV scale so that they can be tested at the colliders such as the LHC. For example, in this model, the dominant decay mode of $W_R$ is not to the type I mode $\ell^\pm\ell^\pm jj$ but rather to $t\bar{b}$  as well as to $\ell$+ missing $E_T$ final states. The LHC limit on $W_R$ for this mode is 4 TeV~\cite{CMS}. Such right handed keV neutrinos may also be accessible at the KATRIN experiment~\cite{werner}. Furthermore, the quark seesaw leads to flavor violating decays of $t$ e.g. $t\to c+g$ etc which can be used to test these models.

\item Since the masses of $\Delta_{L,R}$ are close to $3$ and $1$ TeV respectively in this model, from the flavor structure of the $f$-coupling from neutrino mixings, one expects flavor changing rare decays of the $\tau$ lepton. With $f\sim 0.05$, this is consistent with current limits but this model predicts that they should be observable once the search sensitivity goes down. Clearly, the predictions depend on whether the neutrino mass hierarchy is normal or inverted and we do not go into a detailed discussion of this in this paper.

\item For light $\nu_R$ and low mass $W_R$, there are stringent supernovae 1987A bound~\cite{barmoh} on $W_R$ mass of $\sim 20$ TeV or so. However, it must be pointed out that these bounds are dependent on detailed supernova properties e.g. radial dependence of temperature as well as possible new interactions~\cite{bmy}.
For example, in addition to the interactions in the model, if there are singly charged $SU(2)_{L,R}$ singlet particles $\eta^+_{L,R}$, their interactions could trap the right handed neutrinos~\cite{bmy} and help avoid the above bound.

\end{itemize}

\section{Conclusion}

In this paper, we show how to avoid the lower bound on $W_R$ mass in left-right symmetric models that arises from leptogenesis in  type-I TeV scale models. We consider a quark seesaw version of the left-right model where left-right gauge symmetry breaking is decoupled from leptogenesis by adopting an alternative scenario for lepton number generation. Lepton asymmetry is generated not by the decay of right handed neutrinos as in conventional type I models but rather via the decay of some gauge singlet scalar bosons. The latter decays generate an asymmetry between the lepton number carrying Higgs triplets responsible for type-II seesaw which subsequently decay to leptons via the type-II type coupling and transfer their asymmetry to lepton number. We show how the small triplet VEV required for understanding small neutrino masses is generated in this model via a two step symmetry breaking, thereby alleviating the fine tuning problem associated with the SM implementation of type-II seesaw. Our model predicts keV sterile neutrinos coupling to right handed currents. The strong CP problem is also solved by parity symmetry. Furthermore, the left-right symmetry breaking, the quark seesaw and the type-II seesaw are all at the multi-TeV scale so that they can leave testable signatures at the colliders such as the LHC. For example, in this model, the dominant decay mode of $W_R$ is not to $\ell^\pm\ell^\pm jj$ but rather to $t\bar{b}$ final states as well as to $\ell$+ missing $E_T$. The right handed keV neutrinos predicted by this model may also be accessible at the KATRIN experiment~\cite{werner}.

\textbf{Acknowledgement}: P.H.G. was supported by the National Natural Science Foundation of China under Grant No. 11675100 and the Recruitment Program for Young Professionals under Grant No. 15Z127060004. R.N.M. was supported by the US National Science Foundation under Grant No. PHY1620074.

\end{document}